\documentclass[a4paper]{PoS}

\usepackage{xcolor}

\title{The Cherenkov Telescope Array view of the Galactic Center region}

\ShortTitle{CTA view of the Galactic Center region}

\author{\speaker{Aion Viana}$^{1}$, Sofia Ventura$^{2,3}$, Daniele Gaggero$^{4}$, Dario Grasso$^{2}$, Dmitry Malyshev$^{5}$, Karl Kosack$^{6}$, Stefan Funk$^{5}$, Antonino D'Ai$^{7}$, Rafael Alves Batista$^{8}$, Rebecca Blackwell$^{9}$, Aleksandr Burtovoi$^{10}$, Patrizia A. Caraveo$^{11}$, Masha Chernyakova$^{12}$, Paolo Da Vela$^{2,13}$,  Giovanni De Cesare$^{14}$, Elisabete M. de Gouveia Dal Pino$^{8}$, Emma de O\~{n}a Wilhelmi$^{15}$, Miroslav Filipovic$^{16}$, Michele Fiori$^{10}$, Ignacio A. Minaya$^{17}$, Juan Carlos Rodr\'{i}guez-Ram\'{i}rez$^{8}$, Gavin Rowell$^{9}$, Andrea Rugliancich$^{2}$, Olga Sergijenko$^{18}$, Luca Zampieri$^{10}$ for the CTA Consortium\footnote{for consortium list see PoS(ICRC2019)1177} \\
\llap{$^1$}Instituto de F\'{i}sica de S~{a}o Carlos, Universidade de S\~{a}o Paulo, S\~{a}o Carlos SP, Brazil;\,
\llap{$^2$}INFN Sezione di Pisa,Polo Fibonacci, Pisa, Italy;\,
\llap{$^3$} University of Siena, Siena, Italy; \,
\llap{$^4$}Instituto de F\'{i}sica Te\'{o}rica UAM-CSIC, Madrid, Spain; \,
\llap{$^5$}Erlangen Centre for Astroparticle Physics, Erlangen, Germany ; \,
\llap{$^6$}IRFU, CEA, Universit\'{e} Paris-Saclay, Gif-sur-Yvette, France;\,
\llap{$^7$}INAF - IASF, Palermo, Italy; \,
\llap{$^8$}IAG - Universidade de S\~{a}o Paulo, S\~{a}o Paulo, Brazil; \,
\llap{$^9$}School of Physical Sciences, University of Adelaide, Adelaide, Australia; \,
\llap{$^10$}INAF-OAPD, Padova, Italy; \,
\llap{$^{11}$} INAF - IASF, Milano, Italy; \,
\llap{$^{12}$}School of Physical Sciences, Dublin City University, Dublin, Ireland; \,
\llap{$^{13}$} University of Innsbruck, Innsbruck, Austria; \,
\llap{$^{14}$}INAF - OAS, Bologna, Italy; \,
\llap{$^{15}$}Institute of Space Sciences (ICE/CSIC), Barcelona, Spain; \,\,
\llap{$^{16}$}Western Sydney University, Penrith, Australia; \,
\llap{$^{17}$}University of Liverpool, Oliver Lodge Laboratory, Liverpool, UK; \,
\llap{$^{18}$}Astronomical Observatory of Taras Shevchenko National University of Kyiv, Kyiv, Ukraine \\
E-mail:  \email{aion.viana@ifsc.usp.br}}

\abstract{Among all the high-energy environments of our Galaxy, the Galactic Center (GC) region is definitely the richest. It harbors a large amount of non-thermal emitters, including the closest supermassive black hole, dense molecular clouds, regions with strong star forming activity, multiple supernova remnants and pulsar wind nebulae, arc-like radio structures, as well as the base of what may be large-scale Galactic outflows, possibly related to the Fermi Bubbles. It also contains a strong diffuse TeV gamma-ray emission along the Galactic ridge, with a disputed origin, including the presence of a possible Pevatron, unresolved sources, and an increased relevance of the diffuse sea of cosmic rays. This very rich region will be one of the key targets for the next generation ground-based observatory for gamma-ray astronomy, the Cherenkov Telescope Array (CTA). Here we review the CTA science case for the study of the GC region, and present the planned survey strategy. These observations are simulated and we assess CTA's potential to better characterize the origin and nature of a selection of gamma-ray sources in the region.}

\FullConference{36th International Cosmic Ray Conference -ICRC2019-\\
		July 24th - August 1st, 2019\\
		Madison, WI, U.S.A.}

\begin{document}

\section{Introduction}

The Galactic Center region is the most active region in the Milky Way containing a large amount of sources detected at all wavelengths. It also provides a unique environment in the Galaxy where a recent acceleration and propagation of cosmic-rays has been observed. Gamma-ray emission from this region has been detected for the first time by the 
EGRET satellite (3EG J1746-2852, \cite{egret}) in 1998. Observations with imaging atmospheric Cherenkov telescopes (IACTs) soon after gave rise to the detection of a point-like emission in the very-high-energy (VHE, E > 100 GeV) gamma-ray regime by the Whipple~\cite{veritasgc}, H.E.S.S.~\cite{hessgc} and MAGIC~\cite{magicgc} telescopes. The Fermi-LAT satellite then detected a central source, 1FGL J1745.6-2900, coincident in position with Sgr A*~\cite{fermigc}, though no firm conclusion does exist about the association of the Fermi-LAT source with the VHE gamma-ray source due to the presence of a strong diffuse gamma-ray emission in the same energy range. In order to search for much fainter emission, an analysis of the GC region was made by the H.E.S.S. collaboration~\cite{ridge}, subtracting the best fit model for point-like emission at the position of the central source and the SNR G0.9+0.1. This revealed a ridge of diffuse emission extending along the Galactic plane for about 2$^{\circ}$ in Galactic longitude, which was found to be spatially correlated to giant molecular clouds located in the central molecular
zone (CMZ)~\cite{ridge}. The strong correlation between the morphology of the diffuse gamma-ray emission and the density of molecular clouds indicates the presence of protons propagating in the region and interacting with the cloud material, giving rise to the observed gamma-ray flux via $\pi_0$ decays. However, the origin of these protons is still under debate, including the presence of a possible Pevatron~\cite{pevatron}, unresolved sources~\cite{Hooper:2017rzt}, and an increased relevance of the diffuse sea of cosmic rays towards the inner Galaxy~\cite{Gaggero:2017jts}.

There are currently six firmly detected sources of VHE gamma-rays located within the inner few degrees of the Galaxy: (i) the bright central source, coincident in position with SgrA*; (ii) the composite supernova remnant G 0.9+0.1; (iii) HESS J1745-303, an extended source with complex morphology; (iv) the Galactic Center Ridge; (v) HESS J1741-302, a weak source located on the Galactic Plane; (vi) and the sources HESS J1746-285, MAGIC J1746.4-2853 and VER J1746-289 that are positionally coincident to each other, and that most likely consist of a mix of the emission from the ridge and from a yet unidentified emitter. Despite the intense speculation about the mechanisms responsible for all the observed emissions, no firm conclusion has emerged so far. 

Besides the known sources, a wide variety of possible high-energy emitters, yet undetected in VHE gamma-rays, are also present in the GC region, including young massive stellar clusters such as the Arches and Quintuplet clusters, known SNRs such as Kepler, Sgr C and SNR G1.9+0.3, over 10 PWN candidates, arc-like radio structures, as well as the base of what may be large-scale Galactic outflows, possibly related to the Fermi Bubbles~\cite{fermibb}. Finally, this area is expected to be the brightest source of dark matter annihilations in the gamma-ray sky by several orders of magnitude due to its large DM density and relative proximity to Earth. Even considering a possible signal contamination from other astrophysical sources, it is one of the prime targets to detect the presence of DM particles.

In the next sections we will describe the preparations for the science operations of the Cherenkov Telescope Array (CTA) and how it intends to solve persistent questions about the high-energy emitters in the inner hundreds of parsecs of the Milky Way. In particular, we will show preliminary results of the CTA capabilities to study the physical properties of the two brightest TeV sources in the GC region, namely the GC Central Source and the SNR G0.9+0.1. 

\section{The Galactic Center Key Science Project}

CTA will be sensitive to gamma-rays from  20 GeV to more than 300 TeV. With more than 100 telescopes located in two sites in  the northern (La Palma, Spain) and southern hemispheres (Paranal, Chile), CTA will achieve an improvement by a factor of five to twenty (depending on energy range) in sensitivity with respect to the current major gamma-ray instruments. Based on the current CTA design study, a factor of about ten in effective area, and a factor of two better in hadron rejection and angular resolution are expected. CTA-South will be able to observe the GC region with unprecedented accuracy, in particular because it transits close to zenith in relation to the array. 

\begin{figure}[ht]
	\begin{center}	
		\includegraphics[width=0.49\linewidth]{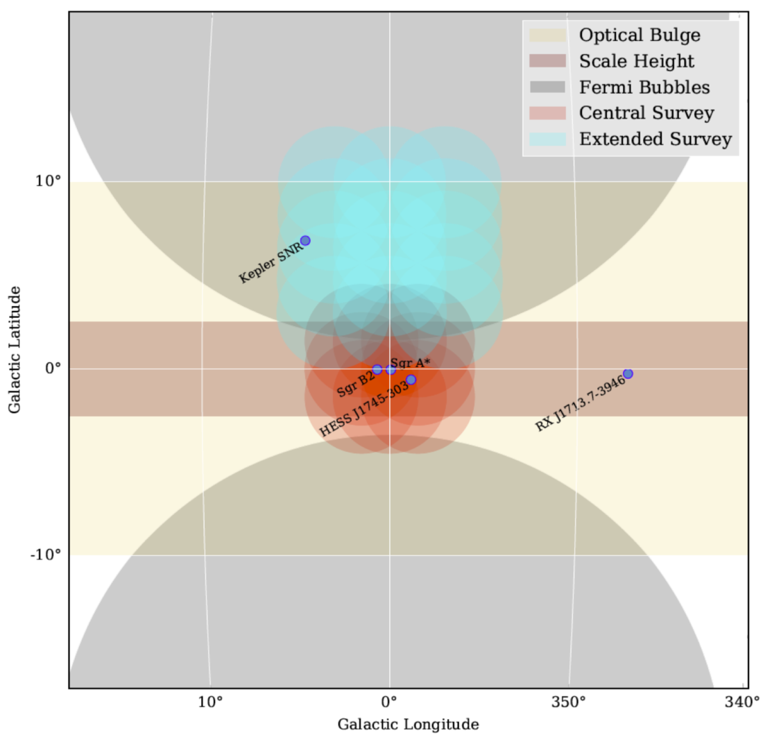}
		\includegraphics[width=0.49\linewidth]{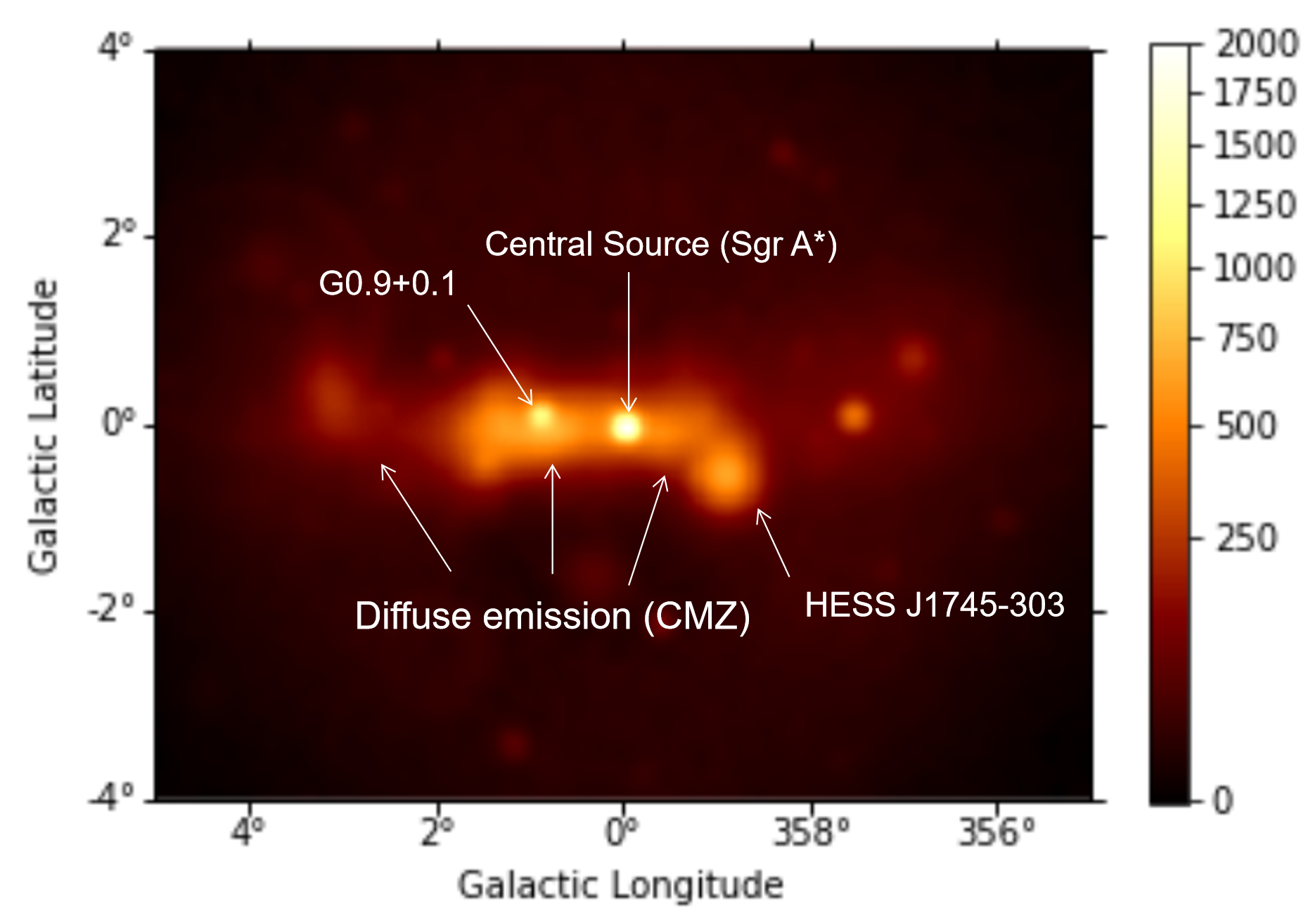}
	\end{center}
	\caption{{\em Left:} A schematic representation of the Galactic Centre KSP. This figure shows one possible observation strategy for CTA. The deep survey region is shown in red, with the Galactic bulge extension shown in cyan (with each circle representing a 6$^{\circ}$ field of view for a typical CTA configuration) (from Ref.~\cite{CTAScience}). {\em Right:} A simulated view of the inner 4$^{\circ}$ of the Galactic Centre region as seen by CTA (excess events above 100 GeV after cosmic-ray background subtraction, with PSF smoothing).}
	\label{fig:gcksp}
\end{figure}

The CTA observatory will be open to proposals from the scientific community. However, during its first decade of operation, approximatively 40\% of the available observing time will be dedicated to a Core Programme. This programme consists of several Key Science Projects (KSPs) - major multi-purpose observations designed to efficiently address a broad ranging of science questions~\cite{CTAScience}. The Galactic Center KSP is comprised of a deep survey of the GC region containing two small survey regions: a deep exposure close in to the central source, centered on Sgr A*, and an extended region starting at the edge of the deep exposure and extending to the edge of the Galactic bulge (see Fig.~\ref{fig:gcksp}). The exact pointing strategy will be optimized once the final field of view and instrument response function of CTA-South is known, and it may also be adjusted to provide, for example, better coverage of interesting features of the Fermi bubbles based on new analyses of GeV data.

This survey of the GC region is considered key science for both scientific and technical reasons. The proposed observations will require a large time commitment spread over several years providing the deepest exposure of the GC region ever produced in the VHE domain. The broad variety of targets and topics is expected to give rise to high-impact scientific results for years to come. As part of this KSP, the CTA Consortium will organize coordinated observations in other wavebands that will permit detailed variability studies and constitute an invaluable legacy data set. 

From the technical point-of-view, the realization of an extremely deep and high-precision survey of the Galactic Centre region is very challenging due to the complexity of this area. Source confusion and the difficulty of modeling the background when there is an expectation of sources that are larger or similar to the instrumental field of view are some of the difficulties. Besides, the necessity to handle strong background optical light that varies by a factor of over ten within the field of view, and the need to push the systematic errors to extremely low levels to produce meaningful limits on models, will require a deep knowledge of the instrument and atmosphere~\cite{CTAScience}.

\section{Data simulations and analysis}

In order to test CTA capabilities to analyse and extract the expected scientific results of the future KSPs, a set of in-depth Monte Carlo simulations of the first three years of observations of CTA was generated in what has been called the first CTA science data challenge (DC-1). High-level science data (observatory products comprised of selected gamma-like events, instrument response tables, and calibration data) of the existing TeV and GeV sky, as well as extrapolations of measured fluxes to higher and lower energies, based on population models, were produced. The event lists were simulated from high-level instrument response functions that correspond to an ideal CTA with good
and stable atmosphere and instrument conditions; only zenith angles of 20 degrees and 40 degrees were simulated. The GC survey was simulated assuming all the 825 (525 + 300) hours of observations distributed on 1671 pointings (following the scheme illustrated in Fig.~\ref{fig:gcksp}), and it included the Galactic Center Central Source (CS), the PWN/SNR G0.9+0.1, HESS J1745-303, and the Galactic Ridge (as part of the Galactic Diffuse Emission). The right panel of Fig.~\ref{fig:gcksp} shows the simulated excess map of gamma-rays above 100 GeV after hadronic background subtraction and smoothed by the CTA point-spread function (PSF). 

\begin{figure}[ht]
	\begin{center}	
		\includegraphics[width=0.48\linewidth]{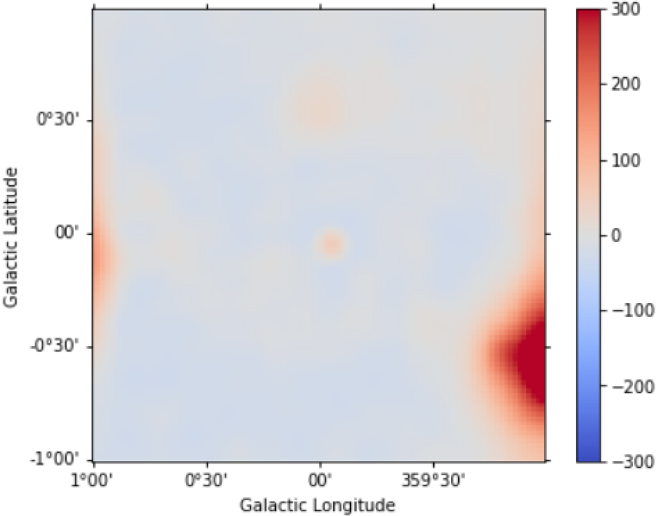}
		\includegraphics[width=0.51\linewidth]{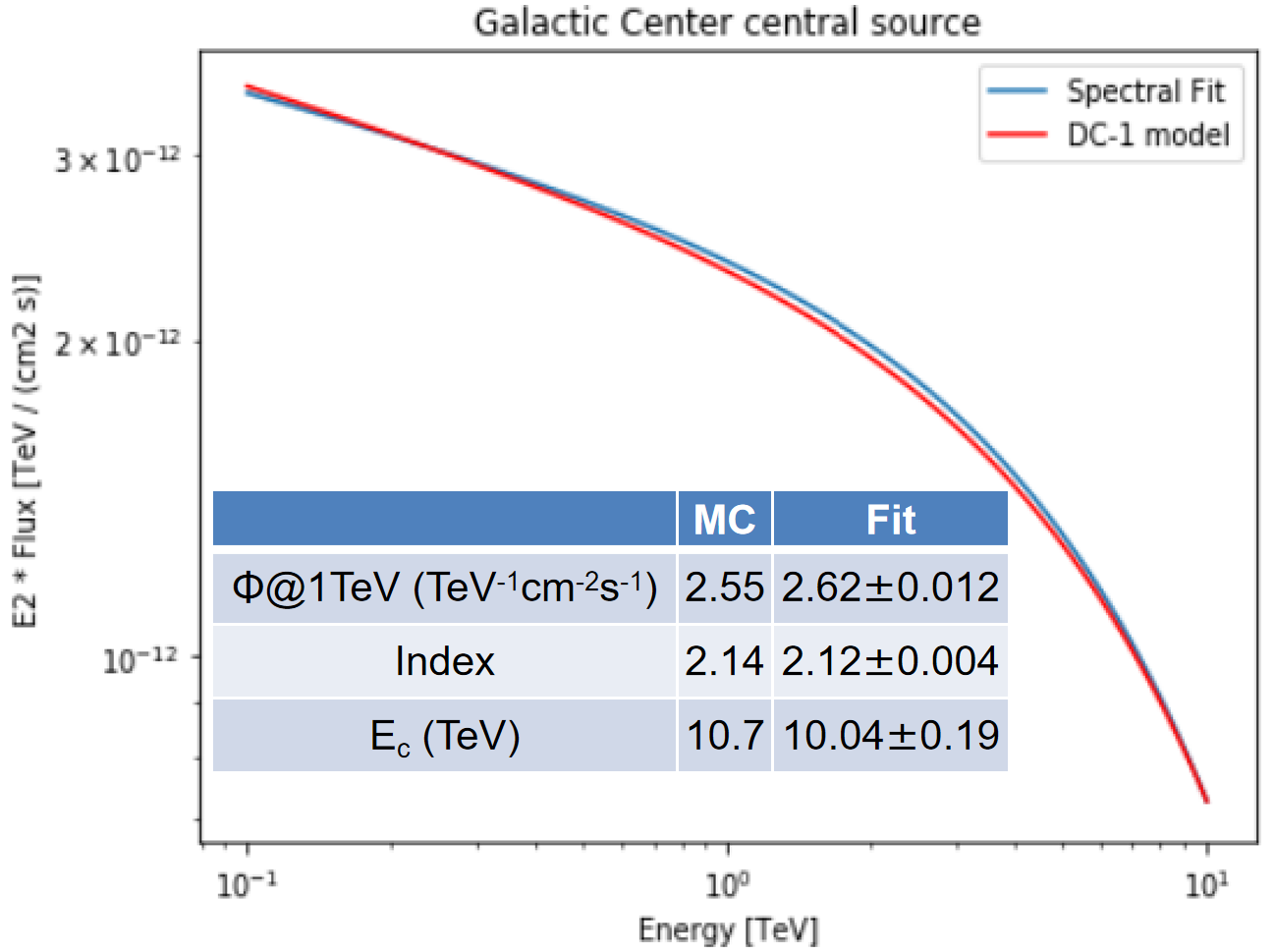}
	\end{center}
	\caption{{\em Left:} Residuals after subtracting the fitted GC central source, G0.9+0.1 and the Galactic Ridge using a 3D-template analysis. {\em Right:} Reconstructed central source spectrum compared to input model. The index, amplitude and energy cut-off are measured with a 0.2\%, 0.5\% and 2\% level of statistical uncertainty, respectively.}
	\label{fig:res}
\end{figure}

The simulated data were analyzed using a 3D template technique, where the gamma-ray sources and the hadronic background are simultaneously fitted under an assumption about their spectral and spatial distribution. The central source and G0.9+0.1 were simulated and fitted as point-like sources, and with a spectrum following a power-law with exponential cut-off and pure power-law shapes, respectively. For the Galactic Ridge emission, predictions from the CR propagation codes Picard~\cite{Kissmann:2014sia}, for inverse-Compton emission, and Dragon~\cite{Vittino:2018ieh} for gas-related emission (bremsstrahlung + $\pi_0$) were used. HESS J1745-303 was simulated assuming a Gaussian morphology with a 0.2$^{\circ}$ radius and power-law spectrum. However, for this initial study, HESS J1745-303 did not participate in the fitting procedure (the spatial and spectral parameters were frozen to the simulation values). On the left panel of Fig.~\ref{fig:res} the residuals after subtraction of the fitted GC central source, G0.9+0.1 and the Galactic Ridge are shown. These sources are successfully reconstructed with high precision. The reconstructed spectrum of the GC central source is compared to the MC true spectrum on the right panel of Fig~\ref{fig:res}. The index, amplitude and energy cut-off are measured with 0.2\%, 0.5\% and 2\% of statistical uncertainty, respectively.

\section{Individual sources}

\subsection{Galactic Center central source}

The nature of the VHE gamma-ray emission at the dynamic centre of the Milky-Way is still unknown. High pointing precision observations by the H.E.S.S. telescope led to the exclusion of Sgr A East supernova
remnant as the main counterpart of the observed VHE emission~\cite{2010MNRAS.402.1877A}, and left Sgr A*~\cite{aharonian2005, liu2006} and the plerion G359.95-0.04 (discovered a lightyear of Sgr A*) \cite{wang2006} as plausible contributing sources for the observed emission. Different non thermal mechanisms have been suggested to explain this source, where a central accelerator of relativistic protons or electrons, or the self-annihilation of dark-matter particles have been suggested to be at the origin of such emission. Dark matter annihilation seems, however, unlikely due to the predicted spectral shape, which would demand a much sharper energy cut-off. 

\begin{figure}[ht]
	\begin{center}	
		\includegraphics[width=0.52\linewidth]{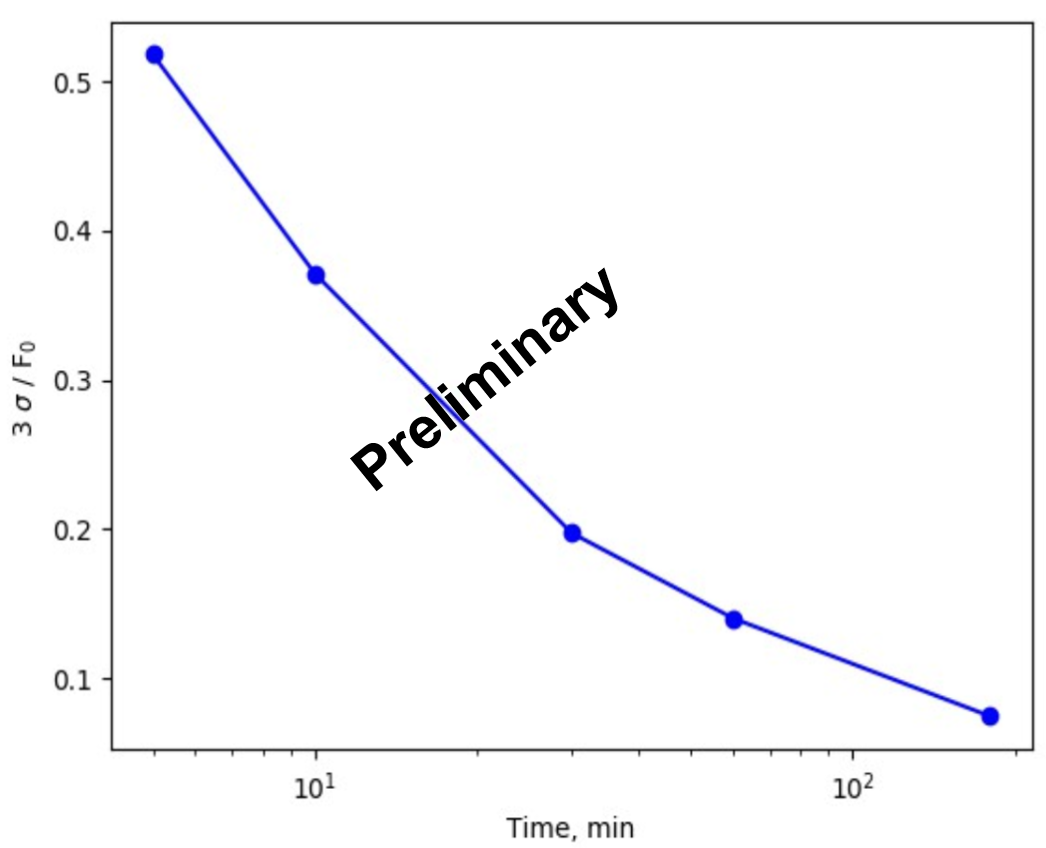}
		\includegraphics[width=0.47\linewidth]{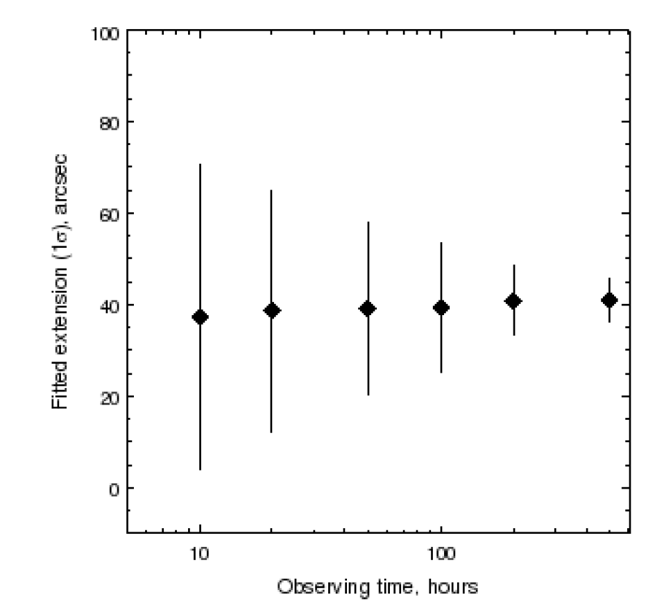}
	\end{center}
	\caption{{\em Left:} Flux increase of the GC central source during a flare for a 3$\sigma$ detection by CTA versus the flare duration. {\em Right:} Fitted size of the central source (assuming a Gaussian shape) made by CTA as a function of observing time~\cite{CTAScience}. }
	\label{fig:sgra}
\end{figure}

Leptonic models include a plerion scenario with Sgr A* as the wind source \cite{liu2006,atoyan2004} , or the pulsar wind nebula G359.95-0.04~\cite{wang2006,hinton2007}. Interestingly, the plerion models are also able to reproduce the IR and X-ray flaring behavior of the central source (Sgr A* emission is known to vary by a factor of a few on 5min to 3h time scales~\cite{2001Natur.413...45B,2003Natur.425..934G,2003AA...407L..17P,2008ApJ...682..373M,2015ApJ...799..123B}). Although no variability has yet been detected in TeV energies~\cite{2009AA...503..817A}, there is still the possibility that the TeV variability is too weak to be detected by current detectors that are only sensitive to changes of flux of roughly a factor of two. In order to estimate CTA detectability of weak flares in VHE gamma rays, MC simulations of the GC central source were performed with different observation time-windows. The flux increase during a flare with respect to the quiescent state in order to have a 3$\sigma$ detection by CTA as function of the flare duration is shown in the left panel of Fig.~\ref{fig:sgra}. In the case of a flare duration of 30 minutes, a flux increase of just 20\% would be detected by CTA at a 3$\sigma$ level.


In most hadronic scenarios, stochastic acceleration of protons interacting with the turbulent magnetic field in the vicinity of Sgr A* could produce an outflowing flux of relativistic protons that diffuse outwards and scatter with the gas of the dense environment close to the GC~\cite{aharonian2005,ballantyne2011,chernyakova2011,linden2012,fatuzzo2012,Rodrigurez-Ramirez:2019fmm}. These protons would interact firstly in the accretion flow region of the SMBH ($\lesssim 0.1$ pc or 2.6 arcseconds), then with the low density gas in the central cavity ($\lesssim 2.9$ pc or 70 arcseconds), subsequently with the high density molecular gas in the circumnuclear ring ($\simeq$ 3 pc or 70 arcseconds), and finally with the local molecular clouds to produce the gamma-ray signal (located within 10 pc of the GC). The morphology of the emission will depend on the CR density distribution during its propagation and on the actual gas distribution. The angular resolution and source localization capabilities of CTA will provide strong constraints on the extension of the central source (see right panel of Fig.~\ref{fig:sgra}). This will help discard many scenarios regarding its nature. For example, if the emission arises from cosmic rays interacting mainly with the circumnuclear ring, then we expect a larger extension (up to 70'' ignoring ballistic propagation) than if the interaction happens in the accretion flow region or if it is produced by the possible counterpart PWN G359, which would give an extension of < 10'' given the magnetic and radiation fields near the black hole~\cite{CTAScience}. Here it is important to point, however, that for this first set of simulations the instrument response functions were not yet optimized for the PSF size. In the case of deep observations, a much better angular resolution can be achieved by selecting only well-reconstructed events, but this study is still underway.






\subsection{SNR G0.9+0.1}




G0.9+0.1 is a composite SNR located in the direction of the Galactic Center and at about the same distance (to be 8.5 kpc). It is characterized by a bright compact radio PWN ($\sim$2' across) surrounded by an extended shell ($\sim$8' diameter). Based on the size of the shell, it has an estimated age of few thousand years. The gamma-ray emission was first detected by the H.E.S.S. telescope~\cite{2005AA...432L..25A} as a point-like source with a spectrum from 0.2-7 TeV following a simple power law with index of 2.3. The emission appears to originate from the PWN core of the remnant due to the point-like morphology, and the lack of strong non-thermal X-ray emission from the shell. The gamma-rays could then be explained as coming from inverse Compton scattering of relativistic electrons. However, both the lack of an extension and a high energy cut-off in the gamma-ray spectrum does not allow for a clear conclusion about the actual astrophysical process in place. 

\begin{figure}[ht]
	\begin{center}	
		\includegraphics[width=0.44\linewidth]{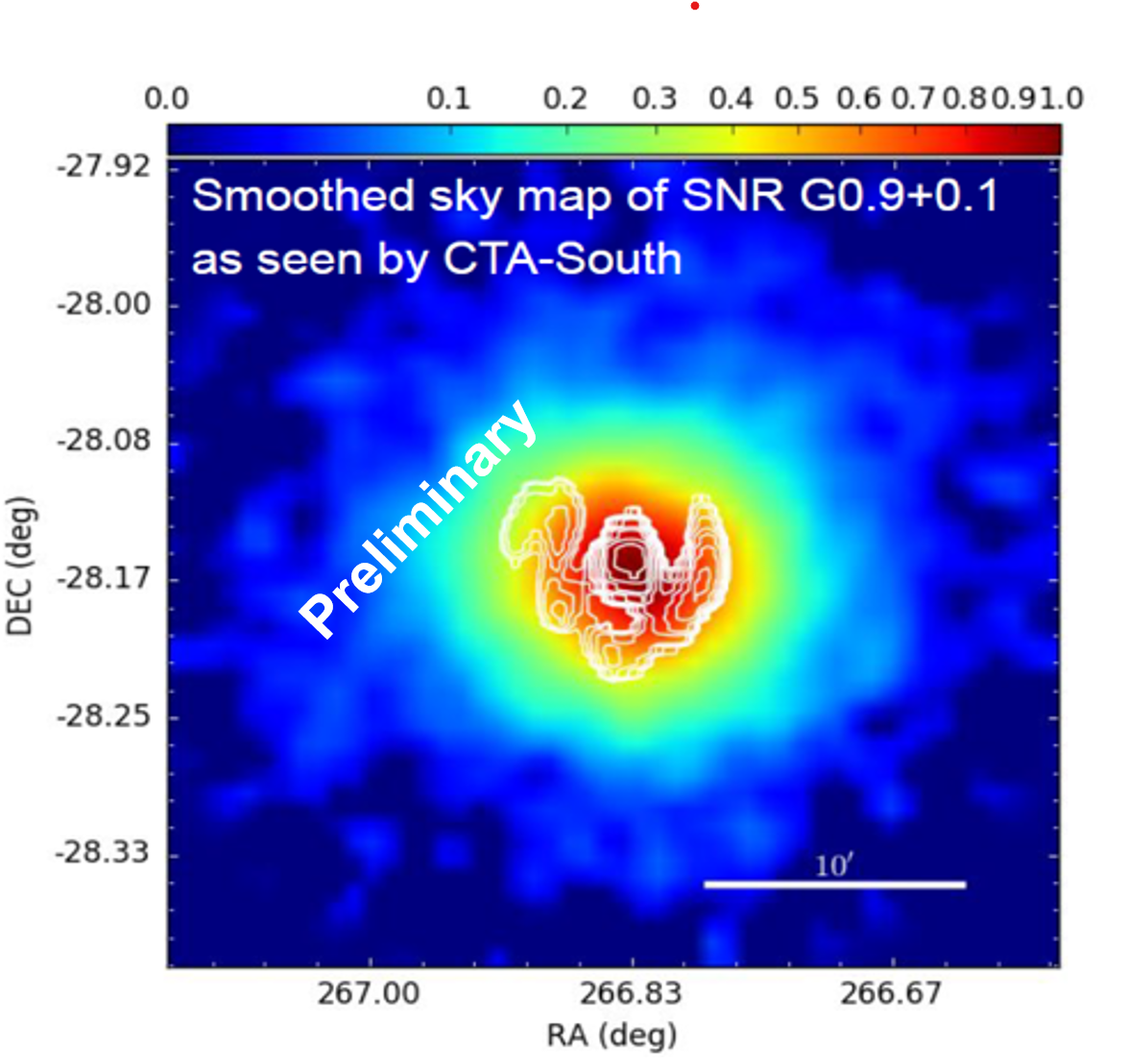}
		\includegraphics[width=0.55\linewidth]{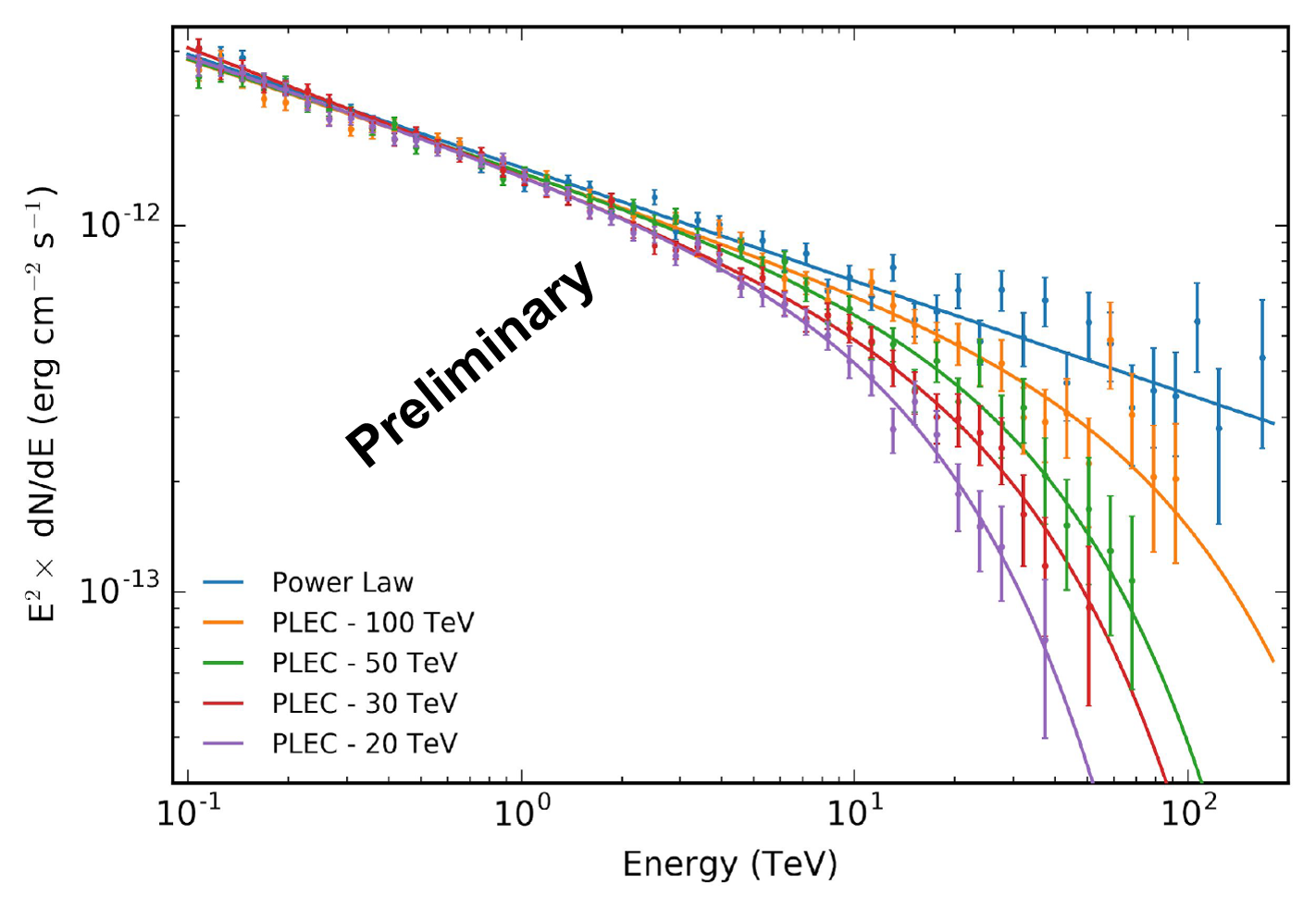}
	\end{center}
	\caption{{\em Left:} Simulated view of SNR G0.9+0.1 as seen by CTA. {\em Right:} Simulated spectra of 30min observation of G0.9+0.1 with different energy cut-offs (from Ref.~\cite{fiori})}
	\label{fig:g09}
\end{figure}

CTA will be able to resolve the VHE emission region if it is larger than $\sim$1.5 arcmin (see left panel of Fig.~\ref{fig:g09}). The simulation of G0.9+0.1 spectrum assuming different high-energy cut-offs and 200 hours of CTA observation is presented in the right panel of Fig.~\ref{fig:g09}. CTA will clearly be capable to distinguish between different spectral models and calculate an energy cut-off up to at least 100 TeV, if present~\cite{fiori}.

\subsection{Conclusions}

In the first few years of CTA operations a large survey of the GC region will take place under the CTA Galactic Center Key Science Project. A preliminary assessment of CTA capabilities to study the two brightest sources in this survey, the GC central source and G0.9+0.1, was presented. It was shown for instance, that CTA will be capable to reconstruct the morphology of these sources with an angular resolution sufficient to image arc-minute scales. It will also be able to detect flares and measure spectral variability on <30 minute scales. Finally, CTA will have sufficient spectral sensitivity and energy coverage to determine the maximum energy reached by accelerated cosmic rays in this region with unparalleled precision. These capabilities will be of utmost importance, for instance, to study the VHE diffuse emission in molecular clouds of the GC region, and to help in the discrimination between different production mechanisms that have been proposed to explain such emission: a Pevatron, unresolved sources and a hard diffusion of the sea of CRs in the inner Galaxy. A complete study that will address the full potential of CTA in the whole GC region, including all known gamma-ray sources and a selection of potentially detectable emitters, as well as the study of systematical uncertainties, is currently in preparation within the CTA Consortium. 

\section*{Acknowledgments}
We gratefully acknowledge financial support from the agencies and organizations listed here: \textit{www.cta-observatory.org/consortium\_acknowledgments}

\end{document}